\documentclass[lettersize,journal]{IEEEtran}
\def\BibTeX{{\rm B\kern-.05em{\sc i\kern-.025em b}\kern-.08em T\kern-.1667em\lower.7ex\hbox{E}\kern-.125emX}}

\usepackage{times}
\usepackage{gensymb}
\usepackage{tabularx}
\usepackage{lipsum}
\usepackage{array}
\usepackage{booktabs}
\usepackage{listings}
\usepackage{amsmath}
 
\usepackage{enumerate}

\usepackage{multirow}

\ifCLASSINFOpdf
 \usepackage[pdftex]{graphicx}
 \else
 \usepackage[dvips]{graphicx}

 \usepackage{amssymb,amsmath}
 
\fi

\usepackage[caption]{subfig}
\usepackage{cite}

\lstdefinestyle{lststyle}{
 captionpos=b, 
 tabsize=2,
 basicstyle=\linespread{0.9}\footnotesize\ttfamily,
}
 
\begin{document}

\title{Improving Remote Patient Monitoring Systems Using a Fog-based IoT Platform with Speech Recognition}

\author{Marc~Jayson~Baucas,~\IEEEmembership{Student Member,~IEEE}, and Petros~Spachos,~\IEEEmembership{Senior Member,~IEEE}
\thanks{This work was supported, in part, by the Natural Sciences and Engineering Research Council (NSERC) of Canada. M. Baucas and P. Spachos are with the School of Engineering, University of Guelph, Guelph, ON, N1G2W1, Canada. (e-mail: baucas@uoguelph.ca; petros@uoguelph.ca) }}

\maketitle
\begin{abstract}
Due to the recent shortage of resources in the healthcare industry, Remote Patient Monitoring (RPM) systems arose to establish a convenient alternative for accessing healthcare services remotely. However, as the usage of this system grows with the increase of patients and sensing devices, data and network management becomes an issue. As a result, wireless architecture challenges in patient privacy, data flow, and service interactability surface that need addressing. We propose a fog-based Internet of Things (IoT) platform to address these issues and reinforce the existing RPM system. The introduced platform can allocate resources to alleviate server overloading and provide an interactive means of monitoring patients through speech recognition. We designed a testbed to simulate and test the platform in terms of accuracy, latency, and throughput. The results show the platform's potential as a viable RPM system for sound-based healthcare services.      
\end{abstract}


\IEEEpeerreviewmaketitle

\section{Introduction}
\IEEEPARstart{A}{s} the global population count rises, and healthcare resource allocation becomes a concern. The recent pandemic has left a mark on taking up physical resources within the medical field and has caused struggles in the healthcare industry. With the rising concern, healthcare turns to novel technologies, such as the Internet of Things (IoT), to examine the potential to alleviate the resource strain~\cite{hc-resource}. 

Healthcare providers have adapted and incorporated IoT networks to address physical resource issues. An IoT is a collection of sensor-equipped devices communicating and sharing data through a shared wireless medium~\cite{fog-iot}. When implemented in a healthcare service, it is deployed and made available remotely to patients and medical centers accessing its data~\cite{hc-iot}. Remote Patient Monitoring (RPM) is an example that uses IoT networks to serve healthcare services.  
Available RPM systems have several advantages, such as the volume of real-time data it collects. However, with the magnitude of information it senses comes issues in privacy, data flow, and service interactability. One of the goals of healthcare monitoring services is to preserve the security of a patient's information~\cite{data-manip}. Wireless communication has vulnerabilities that can result in user information being compromised~\cite{health-iot-privacy}. Also, continuous data transmission results in the exhaustion of the server and creating data traffic~\cite{iomt-data-traffic}. As for interactability, most RPM implementations are passive since they only collect data. The healthcare center is responsible for translating this data into useful information to diagnose the patient. However, this data flow prevents healthcare centers from responding in real-time and makes it difficult to respond to emergencies. 

To further improve the quality of RPM systems, we propose a low-cost fog-IoT platform with speech recognition. In this work, we present the following contributions:
\begin{itemize}
    \item We introduce a fog-IoT-based platform to address issues of RPM systems in patient privacy, data flow, and service interactability. As a result, we provide an improved and profitable network architecture for a more effective and secure means of managing RPM systems.
    \item We design a testbed that simulates our platform and evaluates its feasibility based on latency and throughput against other configurations. The testbed also provides a platform for testing future improvements to RPM systems and addressing its other potential concerns.  
\end{itemize}

The rest of this paper is as follows: Section~\ref{rel} briefly discusses the literature and works related to RPM systems in healthcare using IoT. Section~\ref{sol} describes the design solutions to address the open issues. Section~\ref{prop} presents the proposed platform and its components. Section~\ref{res} discusses the results of the different tests conducted through our platform. Finally, in Section~\ref{conc}, we conclude this work.

\section{Background and Related Works}~\label{rel}
The following discusses the relevant works, challenges, and solutions of IoT-based RPMs in healthcare. 

\subsection{Benefits of IoT-based RPMs in Healthcare}
An RPM system is a healthcare service that equips patients with devices that remotely monitor and measure certain aspects of their health. Through a wireless network, the server receives the data. This data hierarchy allows the remote observation of patients for any anomalies concerning their health. RPM systems reduce patient travel for diagnosis or check-ups~\cite{isolated-rpm}. RPM systems can also cater to patients who cannot be physically present in clinics or needs isolation due to certain illnesses. The elderly and immuno-compromised benefit from this system since travel is not required, while the medical centers monitor them when they are in the comfort of their homes~\cite{elderly-rpm}. 

With the increasing availability of IoT networks, distance is less of a concern due to real-time data transmission. As a result, these are used in tandem with wearable IoT devices to implement RPMs for efficient remote monitoring. An IoT-based RPM implementation in~\cite{smart-rpm} uses smart technology to connect medical centers to patients for automated and remote diagnosis. They use it to integrate medical services into smart devices as a convenient means to present their patients' health status. In~\cite{cemg-rpm}, they propose a capacitive electromyography (EMG) monitoring system to address the limitations of the current contact EMG monitoring system. They use capacitive sensing methodology to design a circuit for RPM services. Another benefit of RPM systems is that it provides an environment that capitalizes on the advancements of IoT sensors and smart technologies to improve the quality of its services that collect and sense physiological data remotely. However, sensors have different caveats due to their design and the advancements in the technologies it uses. Therefore the ongoing RPM innovation trend aims to improve these sensors for a better quality of service (QoS). In~\cite{smart-garment}, they designed a smart garment for respiratory monitoring using notched optical fibre sensing fabrics, which makes the design real-time and non-invasive. In~\cite{optimum}, they proposed an optimum placement for relay nodes in wireless body area networks (WBANs) for better RPM QoS. Due to different body types, they inspected different relay node positions on the human body and provided optimal placements for better transmission range coverage and channel quality.

These implementations highlight the advantage of using RPM systems as it allows the utilization and development of sensor technologies to improve the quality of RPM services. However, each approach focuses on managing specific technologies and their design to address each discrepancy. Although it helps the RPM system that the sensor belongs to, it does not fully address the issues that are common to all. Therefore, we aim to supplement this cause by proposing a platform to improve the challenges of RPM systems on data security, data flow, and service interactability.

\subsection{Challenges of IoT-based RPMs in Healthcare}
The following discusses the issues in patient privacy, data flow, and service interactability in current IoT-based RPMs.

\subsubsection{Data privacy} The first issue of RPMs is the safety of its patient's information. Patient privacy is essential for any network-based online healthcare service~\cite{health-iot-privacy}. Patient confidentiality has always been a strictly valued policy. However, monitoring devices collect the patient's information through the network. This data path adds a transmission layer vulnerable to security attacks. Also, due to the nature of wireless networks, data transmissions are over long distances. This further exposes the data to more security attacks. Also, data manipulation can result in severe healthcare problems and wrong diagnoses, which could lead to issues as dire as the loss of life~\cite{data-manip}. Healthcare services are responsible for addressing all these security concerns. As a result, people will remain skeptical of RPM systems without a secure means of maintaining data privacy. Therefore, in an era where network security is a big concern, RPMs must focus on fortifying their systems to keep patient data confidential and secure~\cite{rpm-security}. 

\subsubsection{IoT network data flow} This issue points to the ability of the network to maintain a consistent data flow. Health care services that use RPM services must receive patient data at a reliable rate to be effective~\cite{remote-data}. As a result, RPMs incorporate wearable IoT devices that monitor their patients. These devices collect patient data and send them through the wireless network. Patient monitoring sessions are for long-term operations. This design ensures that the RPM system can function and analyze data continuously for effective diagnosis and health status analysis. As a result, wearable IoT devices used to collect patient data remain on for long periods of time~\cite{energy-wearable}. Also, with a continuous influx of data from multiple devices, the monitoring system could run into issues with central server overloading~\cite{server-overload}. Therefore, a more reliable and adaptive medium for long-term data transport is needed. 

\subsubsection{RPM service interactability} Another issue is the service and patient interactability within the RPM system. RPMs lack a proper means of allowing the healthcare service to interact with its patients~\cite{service-usability}. Raw data is not enough to fully diagnose a patient. That is why some patients are still required to schedule in-person checkups. However, there are events when the patient is not able to comply. Therefore, a more interactive and usable system that caters to a wide demographic will be more convenient for the patient. Also, it could free local healthcare resources and have them redirected to other demands. However, without proper interactability, emergencies are harder to handle within RPMs. As a result, certain situations will be harder to assess while the healthcare centers will have difficulties providing support~\cite{service-support}.

\subsubsection{Other issues with IoT in Healthcare} Aside from what we highlighted, other concerns are proper device utilization, power consumption, and many more. RPM systems require devices that maximize the sensing technologies that they use. If the device or even the system governing it does not have proper energy-efficient resource allocation, the system will have high operation costs~\cite{utilization}. As a result, the service can be less effective in collecting and monitoring patients. Also, power consumption is significant to ensure the sustainability of a system~\cite{energy}. Energy efficiency can help reduce the need for frequent maintenance, charging or battery changes for IoT sensors. Although these are reasonable concerns, we focused more on designing a platform that focuses on the overarching service administration and the network architecture that manages it. Issues in proper device utilization and power consumption deal more with the individual IoT devices and cloud servers used by the RPM service. Our approach targets aspects of the RPM service that deals with managing the IoT network and its data and improving it. 

\subsection{Technologies for Addressing IoT-based RPM Challenges}
The following presents different technologies used in RPM to address these issues and discusses how our proposed design differs from each approach. 

\subsubsection{Wearable IoT devices} 
Many recent works implement a portable RPM system using wearable IoT devices to extend healthcare services remotely. Patients wear these monitoring devices that transmit their vitals to a configured healthcare centre. In~\cite{rigid-flex}, they designed a rigid-flex wearable health monitoring sensor patch for different healthcare services. They use this device as a sensor and gateway combination for data collection and transmission. In~\cite{tele-ecg}, they designed a portable wearable Tele-ECG monitoring system. It uses a redesigned singlet for electrocardiogram (ECG) reading and a smartphone to store and transmit the collected data. Current RPM services focus on designing specialized wearable IoT devices to cater to what it is monitoring~\cite{review-wearable}. 

Our proposed approach is different because we aim to focus more on the security of the system and a design that can operate using low-cost technologies. In addressing the security issue, we cannot reduce the reliance of the monitoring system on the quality of transmitted data. However, we can reduce the chances of compromised data by making a more reactionary platform. Our design improves the current systems by creating a more secure intermediary medium that only transmits data when needed. It allows us to protect the patient's privacy by focusing only on data vital to the user's health. 
    
\subsubsection{Smart systems}
Current implementations of RPM systems take advantage of the architecture of smart technologies to reinforce their design. In~\cite{smart-dfu}, they designed a cloud-based deep-learning platform for the remote detection of diabetic foot ulcers. They used smartphones to collect image data from patients that go to the cloud for analysis, detection, and response. In~\cite{smart-virtu}, they designed a remote monitoring system for the physical rehabilitation of stroke patients. They used cloud IoT architecture partnered with custom smart technology to create a virtual space for patients to exercise and for the clinic to monitor them remotely. These designs used smart architectures to develop centralized medical data and analysis networks. 

We aim to improve the QoS by addressing patient-related issues using low-cost and distributed approaches. Current smart designs revolve around collecting and depositing data to a central server for analysis through the cloud~\cite{smart-rpm}. As a result, most processes are at the center of the network. We focus on creating an alternative that is distributive. These related works on RPM systems use centralized smart technology that only uses its devices as data-sensing components. We propose a platform that reallocates these smart capabilities to the end devices for faster emergency response. Although these technologies can address some issues, their scope cannot cover them. Therefore, we propose an RPM system where the healthcare centre monitors patients using a fog-based IoT network. It combines the beneficial aspects of these technologies for improved services.

\section{Proposed Fog-IoT system}~\label{sol}
An IoT network is a collection of wireless devices communicating within an established hierarchy of data paths~\cite{fog-cloud}. It has a server, a gateway, and a group of IoT devices. A fog-based IoT network is an architecture variant that replaces the gateway with a fog device. It is a technology capable of more complex processing tasks. Also, the fog uses other devices, not the cloud server, to carry out tasks and processes. Local servers or fog devices allow the network to reallocate data packets and redirect data paths to reduce the strain going to the central cloud server~\cite{fog-offload2}. It also extends the network where distance becomes less of a factor, such as in RPM systems. Without requiring the patient to be in the clinic for every checkup, travelling costs and strains are less of a concern. This convenience reduces the need for physical spaces in health centres, which lessens overloading issues~\cite{fog-offload}. However, only adding a fog device is not enough to address the cited issues. Therefore, the following approaches are parts of our proposed solutions. 

\subsubsection{Active data filtering via regulated and adaptive fog servers} 
A significant concern when using RPM systems is patient privacy. We propose a security layer created through a data filter to reinforce this security and prevent potential vulnerabilities. This filter will serve as a gate similar to a firewall that only allows certain types of data to flow through. In turn, data that travel through the network can be better regulated. Our platform adds a fog device between the end devices and the server. It can control the data before it leaves the local network and enters the wider area, being subject to more security attacks. We can also use the fog device to package and pre-process the patient's data into sizes that cause less traffic in the network. Therefore, we propose an adaptive security layer that uses the fog to regulate incoming and outgoing patient data.

\subsubsection{Improved data path control via fog server process reallocation}
Another issue is the data flow. With data transmitted from wearable IoT devices continuously, RPM servers can be subject to overloading and disruptions in the data flow. Any form of delay to the data flow can cause the quality of the service to diminish. As a result, the response time of RPMs towards emergencies and diagnosis will become lacking. The network can reallocate processes from the cloud server to the fog through a fog device. Also, incorporating fog devices into the network can reduce the number of data entry points to the server. This shift in the network hierarchy can reduce the overall amount of data transmitted directly to the server. Partnered with the data filter, controlling the data path can help alleviate network strains due to potential overloading.

\subsubsection{Smart interactive system via an RPM design using speech recognition}
The last issue was service interactability. RPM systems provide convenience to the patient in terms of distance. However, we need more interactability to improve the experience for the patients. We propose using a built-in speech recognition system to allow the patient to interact with the service. Also, we integrated a controlled surveillance system using a built-in camera. Each can help the healthcare centre monitor patients more closely without third-party mediums. We introduce a platform that uses these two interactive mediums to aid the system in connecting the patient with the centre without concerns about distance, clinical resources, and space. Instead of conducting checkups through phone calls or calling in the patients for physical checkups, our platform proposes an interactive means of remote checkups using the RPM system. This design creates a low-cost RPM system that allows patients unable to travel for checkups to seek medical help. Also, it lets clinics with capacity issues take in patients without pushing them into waiting lists.

\section{System platform}\label{prop}
This section covers the different implementation and design aspects of our proposed platform. 

\begin{figure}[t!]
\centerline{\includegraphics[width=0.55\columnwidth]{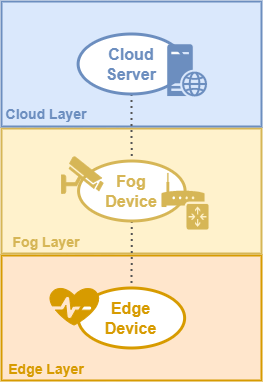}}
    \caption{Network hierarchy of proposed platform.}
    \label{network}
\end{figure} 

\subsection{Overview}
Our proposal is an RPM platform using a fog-based IoT network. Within this network is a fog device design that filters patient data before it goes to the cloud server. Also, it is an interactive tool to allow the healthcare service to conduct remote checkups with the patient. We simulated this smart interactive system using a data-sensing model based on a neural network. We trained it to sense through speech recognition as its core interactive tool. A listening device carries out the sound collection to record the patient's responses. Also, we incorporated controlled video surveillance for visual interactions between the clinic and the patients. With the data filter, only permitted data will leave the local network. 

\begin{figure}[t!]
 \centerline{\includegraphics[width=0.75\columnwidth]{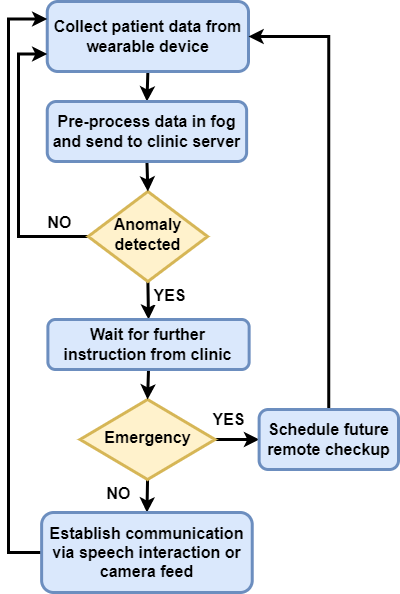}}
    \caption{General design flow.}
    \label{rpm-gen}
\end{figure}

The data flow in the network hierarchy starts with the end devices. The system will use the device to transmit patient information to the fog device. Also, it will be responsible for recording the patients' responses as they interact with the system. This device allows the healthcare centre to communicate with the patient in case of checkups. Next, the fog server will process the data before it goes to the server. When a video feed of the patient is required, the controlled surveillance will turn on the camera. If it is not needed, the platform is in sleep mode except for the wearable IoT device that monitors the patient. However, if the patient cannot respond, this system can also be used as a reactionary measure. If the system detects an anomaly from the data transmitted by the monitoring devices, the clinic can attempt to interact with the patient. This reaction allows the clinic to react accordingly to different situations as they see fit. Also, the data filter remains online. Only essential information will be sent from the fog to the cloud server as agreed upon by the patient and the healthcare centre. Also, this allows it to regulate the data traffic. This agreement enables proper confidentiality and security for the patient. Lastly, the cloud server is the general storage of filtered information and the final say in RPM service decisions. A diagram of the logic flow behind the system is in Fig.~\ref{rpm-gen}. Also, a figure showing the network hierarchy of the proposed platform is in Fig.~\ref{network}.

\subsection{Components}
Given the overview, our aim for a low-cost simulation contains the following hardware and software components. We based these parts presented on the IoT network devices used. 

\subsubsection{Fog device}
This device is the local server. It pre-processes the sound and houses the speech recognition model. Also, it is in charge of filtering the patient data before it goes to the cloud. For the fog device, we used an NVIDIA Jetson TX1 developer kit. It is a development board for prototyping computational tasks and complex software-defined processes. We selected this board due to its built-in graphics processing unit (GPU) and camera peripherals. As a result, it provides a programmable medium optimized for image recording and complex data processing. The Jetson came pre-installed with an Ubuntu 16.04 operating system. We used this built-in GPU as the main sound and general data processing device. It runs the speech recognition model while implementing the data filter for the fog.

The speech recognition model contains feature maps that use a Convolutional Neural Network (CNN). We selected this neural network due to its advantages in modeling and representing acoustic and other sound-based applications~\cite{cnn-sound}. The programming language used to create the CNN model was Python. Tensorflow and Keras were the libraries used to program the training scripts. Our CNN design uses four layers; Convolutional, Pooling, Normalization and Dense. We stored the first three of these layers as blocks. Three identical copies of these blocks are chained to complete a sequence. As the input passes through this sequence, the code flattens the outcome before the dense layer classifies it. We tested this model on previous work with two terms; ``yes" and ``no"~\cite{baucas}. We made this design choice to create a test model for simple interaction. This work differs from it as we decided to utilize the whole Speech Command dataset~\cite{warden2018speech} with all 30 labels totaling 65000 samples to increase the model's complexity. As a result, our test can better represent the speech recognition model as we test it with a collection of more words.  





Aside from speech recognition, the fog device is also in charge of managing the camera. The Jetson has a built-in camera module that is fully programmable. The live capture of the camera can be obtained and streamed to the cloud server upon request. This module is programmed to be controlled by the speech recognition unit and cloud server. The stream is 720p by default. The camera provides information when a face-to-face checkup patient is needed or in an emergency. These functions allow a more interactive RPM system that allows remote checkups. We can obtain interactive data while protecting the patients' privacy partnered with the data filter.



\subsubsection{Cloud server}
The last device in the hierarchy is the cloud server. It stimulates the healthcare centre that receives the processes and filtered data from the fog server. We used a wireless socket to bridge the cloud and the fog. Also, the server that we used in our simulation was a standard PC. It is running on an AMD Ryzen 7 3750H at 2.3 GHz. Also, it uses a 64-bit operating system with an Ubuntu 16.04 image on Windows 10. We stored the data received from the fog in an arbitrary database.

\subsubsection{End devices}
These devices monitor the patient and transmit their data and responses to the fog device. The platform uses Raspberry Pi 3B to simulate the monitoring devices. We selected these devices due to their programmability and modularity. Also, it allows rapid prototyping to represent multiple patients. We loaded each Pi with a Raspbian operating system. We used Pis to simulate the wearable IoT devices that transmitted patient data. This data would then go to the fog device for further pre-processing. Each Pi was also equipped with a sound recording device to sense the responses from the patient. We used the built-in microphone capabilities of an STM32 NUCLEO-64 board with an attached X-NUCLEO-CCA02M1 extension. We chose this device due to its low-power state and programmability. It also includes optimal recording streams and formats that the Pi can use. These microphones were attached via the USB hub and programmed and controlled by Python scripts. The library used for the program is called PyAudio. It allows the creation of recording streams that fit our platform. The configured settings of the recording device were a sampling rate of 16~kHz and 16~bits of resolution. For benchmarking, a table summarizing the parameters of the technologies we used to build the proposed platform is in Table~\ref{spec-table}.



\begin{table}[t!]
\caption{Specifications of the technologies used for designing the proposed Fog-IoT system.}
\label{spec-table}
\fontsize{10}{12}\selectfont
\centering
{
\begin{tabular}{|p{1.7cm}|p{1.8cm}|p{2.0cm}|p{1.6cm}|}\hline
\textbf{Parameter}  & \textbf{Cloud} & \textbf{Fog} &\textbf{Edge} \\\hline
\textbf{Technology}  & ASUS TUF Laptop  & NVIDIA Jetson TX1 & Raspberry Pi 3B \\\hline
\textbf{Processor}  & AMD Ryzen 7 3750H & Quad-Core ARM Cortex-A57 & Quad-Core ARMv8 \\\hline
\textbf{Speed}  & 2.3 GHz & 1.4 GHz & 1.2 GHz \\\hline
\textbf{OS}  & Windows 10  & Ubuntu 16.04 & Raspbian \\\hline
\textbf{Language}  & Python 3.6 & Python 3.6 & Python 3.6 \\\hline
\textbf{RAM} & 8 GB & 4 GB & 1 GB \\
\hline
\end{tabular}
}
\end{table}

\section{Experimental Results} \label{res}
We conducted several experiments on the proposed platform to test the effectiveness of its design based on the highlighted metrics. We then discuss the results to cite any strengths and evaluate our design.

\subsection{Testbed Overview}
After setting up the components according to what we presented in the previous section, we created a low-cost testbed to test the platform's feasibility. We used it to conduct tests within a controlled workspace. These components simulate the data path of the RPM service. Also, we chose a low-cost approach to establish a minimum benchmark for implementing our design and prove the platform's feasibility even with low-end requirements. First, we set up the Pis, the Jetsen TX1 development board containing the fog server, and the laptop running the cloud services within a room with access to the wireless network. Next, we ensured all components have connected to the same WiFi network. Then, we set up each device to connect to the same wireless socket through a shared port. After confirming that each device can transmit data with one another through socket communication, we arranged them according to the network configurations that we will use for our experiments. 

There are two network configurations that we will use to evaluate our platform. One will be cloud-based, while the other fog-based. Our proposed design is fog-based. It treats the gateway as the intermediary device for reallocated processes. These processes are the ones used for speech processing and data filtering. As a result, the gateway becomes the fog device, becoming another medium for carrying out service processes. The cloud-based configuration is the standard network arrangement. The cloud server will be in charge of managing all service processes. Unlike the fog-based setup, the gateway can only reroute data if it is cloud-based.

We selected metrics when testing the platform to show its effectiveness as a solution to the cited issues in this work. The first metric is the accuracy of the speech recognition model. We selected it to gauge the ability of the model to create an interactive RPM service using speech recognition. An accurate means of classifying recorded speech is needed to improve the interactability of RPM systems. The neural network trained the classification model. Also, we tested it using the  Speech Command dataset~\cite{warden2018speech}. Data latency and throughput between the devices within the wireless network were the selected metric. We tested against a standard cloud setup to prove that a fog-IoT network is ideal for this platform. For the latency, we measured the delay between the initial data transfer to the cloud and the completion of its transmission. As for the throughput, we counted the number of packets the server receives under a fixed amount of time.

\begin{figure}[t!]
\centerline{\includegraphics[width=0.95\columnwidth]{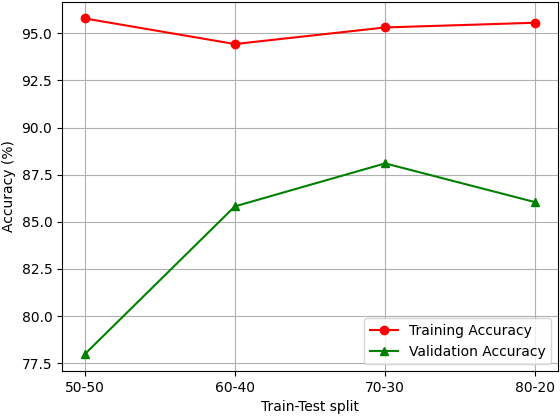}}
    \caption{Model accuracy under varying dataset splits.}
    \label{rpm-acc}
\end{figure}

\subsection{Training the Network}
The first test was for the accuracy of the speech recognition model. We conducted this test to ensure that the interactive design of the platform is at a level where adequate interactions with the patient are possible. A competent speech recognition model partnered with a decent amount of words to give basic instructions can fulfill our goal of a smart interactive service. The complexity of the model increased to create a more adaptive speech recognition system. We tested varying dataset splits to determine which training parameters yield the best accuracy. The plot showing the parametric sweep results is in Fig.~\ref{rpm-acc}. Based on the results, the model peaked accuracies at different dataset splits. Training peaks at 50-50 with 95.79\% and validation at 70-30 with 88.10\%. However, all the training accuracies are around 95\%, while the validating accuracies are more spread out. Therefore, the argument for the best split falls more on the validation accuracy results. Looking at the plot, the 70-30 yields the best overall accuracy results since it has the highest validating accuracy and a training accuracy of 95.31\%. This value is only 0.48\% off the highest. Also, the results from a previous iteration of our work in~\cite{baucas} were 97\%. It makes our results respectable even after increasing its complexity using all 30 words of the whole dataset. Therefore, we can confidently say that we were able to maximize the00 model's accuracy.    

\subsection{Latency}
We also tested the platform's behaviour in terms of latency in addition to the model accuracy. We compared the proposed platform against a standard cloud configuration. This arrangement takes away the fog device, which makes the end devices directly send the data to the cloud server. As a result, the job of processing the sound data and information filtering is at the cloud server. However, the Jetson board still manages the camera. Therefore, the server must treat it as a peripheral for checkups and emergencies.

\begin{figure}[t!]
\centerline{ \includegraphics[width=0.95\columnwidth]{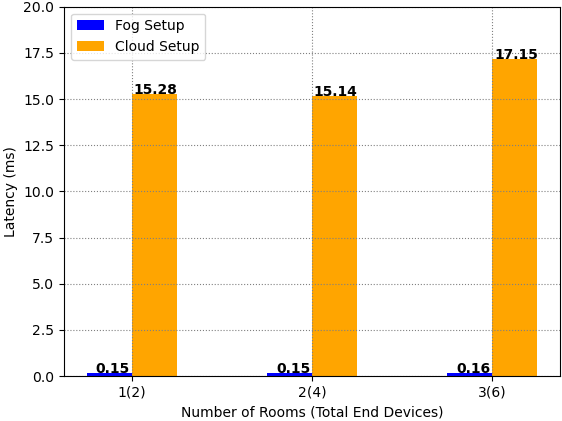}}
    \caption{Average latency of Fog and Cloud setups within 30 iterations.}
    \label{latency} 
\end{figure}

We chose the round trip latency to fully measure the ability of the server to respond to the data sent from the end devices. We experimented by increasing the number of patients per room attempting to communicate with the server. The cloud setup will have rooms with multiple patients. However, each patient's device will independently send data to the server. Meanwhile, the proposed configuration will have the Jetson board receive the data from the patients and send them to the server. We conducted the tests by measuring the average latency when we increased the number of end devices. We grouped the end devices in rooms to show the ability of the platform to reallocate server loads. Arbitrarily, each room contains two end devices. We cycled the measurement of the latency for 30 iterations for each setup. The plot showing the results that simulate up to three rooms are in Fig.~\ref{latency}. Based on the results, we can observe that the cloud setup yields significantly higher latencies on all iterations than the fog setup. The number of endpoints created by the increase in devices increased impacted the cloud setup more than the fog. The cloud's performance in responding to incoming data decreased in quality as more sources connected to the network and transmitted information. Compared to the results of the fog setup, this shows the sensitivity of the network's latency to the number of endpoints it has. 

\subsection{Throughput}
We evaluated the platform's throughput by measuring the number of packets each configuration can do under a given time. Since the platform will manage large sound files from the Pis, it will segment them into data packets. Also, this data needs transmission to be complete so the server can process the data accurately. Therefore, high packet throughput is crucial for efficiency. 

\begin{figure}[t!]
\centerline{\includegraphics[width=.95\columnwidth]{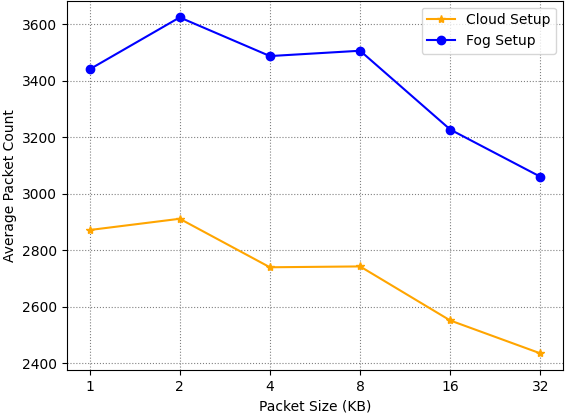}}
    \caption{Average packet count of Fog and Cloud setups within 1 minute of data transmission.}
    \label{throughput} 
\end{figure}

First, we chose a two-room arrangement. So for the cloud, four Pis will send data directly to the cloud server. As for the fog, there will be two and will represent fog devices sending the data to the cloud server. Each iteration of the experiment had the data source send data to the server for one minute. Next, we counted the average number of packets transmitted by each configuration. Also, we increased its size for each measurement from 1, 2, 4, 8, 16, and 32 KBs to further observe the performance of the configurations against growing data transmissions. The plot showing the results of the throughput experiment is in Fig.~\ref{throughput}. Based on the results, we can observe that the fog setup yields an overall higher throughput for packet transmissions. Overall, the average throughput of the experiments decreased as the packet sizes increased. It shows that the data size sent affects the responsiveness of the IoT network as the window of receiving information is smaller as bigger packets enter the server. This sensitivity to packet size becomes more apparent for the cloud setup as the window for taking in data is smaller, with only the cloud server managing all the incoming information. As for the fog, with multiple local servers, we have a bigger window of receiving data with more accessible servers to process the incoming packets. As a result, the downward trend caused by increasing the packet size has less impact on the fog setup throughput than the cloud. 

\subsection{Results and Discussion}
Based on the observations from the experiments, these findings show the platform's capabilities. In terms of accuracy, the tests highlight the potential and progress of the model in handling complex datasets. The classification model shows how it can effectively detect simple keywords we used in training with a testing accuracy of 95.31\% and peak validation accuracy of 88.10\%. As a result, simple decisions and directions require basic information from the patient. 

As for the latency tests, the cloud setup results are significantly higher than the proposed platform. These results can be due to the data traffic caused by having every end device transmit to the server. At the same time, the latency decreased by reallocating some processes to the fog, making the network easier to manage. As for the throughput tests, we can attribute the downward trend to how increasing the packet size reduces the overall possible packet count of the network given a uniform time frame. These results show that the proposed fog-based platform is better at efficiently transmitting data. Also, better throughput depicts a network more capable of managing its data. As for interactability, it has real-time patient-server interactions with better packet throughputs. 

Although the testbed proved the feasibility of the platform, there are limitations to this design. Since we focused on only addressing data privacy, data flow, and service interactability, our design choices were more on reinforcing network and service management. As a result, the platform can only investigate the overall architecture of the network and measure related metrics. Therefore, our aim for future iterations and a new research direction is to improve the platform's ability to integrate different IoT devices. By creating a universal design, we can cater to a more diverse selection of IoT devices and RPM services. Also, we can reduce the impact of the issues in IoT devices, such as operational costs and proper device utilization, making each service more effective for its patients and providers.   

The success in simulating the platform shows that our proposed Fog-IoT platform can operate an RPM system in managing speech data. The proposed platform design creates a data path that enables sensing, transmitting, and storing patient data over devices classified as part of an RPM service through the testbed. Therefore, it can be used for better interactability between the patient and the physician, even though they are remotely apart. Also, it provides a secure and effective RPM service administrator through the blockchain and the Fog-IoT network.

\section{Conclusion} \label{conc}
Current RPM systems run into issues with patient privacy, data flow, and service interactability. As a result, it is not an ideal option for those who are seeking treatment. However, RPMs are the only option for patients who cannot travel to clinics. Also, RPMs may have potential due to the limited space in medical centres. Addressing the issues in RPMs can better help it as a possible option for future treatments. We propose a fog-IoT-based platform that serves as a data filter and interactive model. We used fog technology to reallocate processes from the server to reduce network latency and improve its overall throughput while it uses the data filter to protect the patient's privacy. We used speech recognition and controlled surveillance to upgrade the interactability of the RPM system.

We tested the platform for the accuracy of its speech recognition model. A dataset split of 70-30 with an epoch count of 10 resulted in a training accuracy of 95.31\% and a validation accuracy of 88.10\%. We then tested it against a standard cloud arrangement using latency and throughput as metrics. The latency results show our proposal's significance in regulating data flow with lower delays. As for the throughput, the results show our proposal catering to better real-time interactability with higher packet transmission counts. Overall, the proposed platform has the potential to improve RPM systems.

\bibliographystyle{IEEEbib}
\bibliography{IEEEabrv,rpmbib}

\begin{thebibliography}{10}

\bibitem{hc-resource}
Omar Maki, Mays Alshaikhli, Murat Gunduz, Khalid~Kamal Naji, and Mahmoud
  Abdulwahed,
\newblock ``Development of digitalization road map for healthcare facility
  management,''
\newblock {\em IEEE Access}, vol. 10, pp. 14450--14462, 2022.

\bibitem{fog-iot}
Md~Whaiduzzaman, Md. Julkar~Nayeen Mahi, Alistair Barros, Md.~Ibrahim Khalil,
  Colin Fidge, and Rajkumar Buyya,
\newblock ``Bfim: Performance measurement of a blockchain based hierarchical
  tree layered fog-iot microservice architecture,''
\newblock {\em IEEE Access}, vol. 9, pp. 106655--106674, 2021.

\bibitem{hc-iot}
Mohammad~Nuruzzaman Bhuiyan, Md~Mahbubur Rahman, Md~Masum Billah, and Dipanita
  Saha,
\newblock ``Internet of things (iot): A review of its enabling technologies in
  healthcare applications, standards protocols, security, and market
  opportunities,''
\newblock {\em IEEE Internet of Things Journal}, vol. 8, no. 13, pp.
  10474--10498, 2021.

\bibitem{data-manip}
Gagangeet~Singh Aujla and Anish Jindal,
\newblock ``A decoupled blockchain approach for edge-envisioned iot-based
  healthcare monitoring,''
\newblock {\em IEEE Journal on Selected Areas in Communications}, vol. 39, no.
  2, pp. 491--499, 2021.

\bibitem{health-iot-privacy}
Ke~Wang, Chien-Ming Chen, Zhuoyu Tie, Mohammad Shojafar, Sachin Kumar, and Saru
  Kumari,
\newblock ``Forward privacy preservation in iot-enabled healthcare systems,''
\newblock {\em IEEE Transactions on Industrial Informatics}, vol. 18, no. 3,
  pp. 1991--1999, 2022.

\bibitem{iomt-data-traffic}
S.~Gopikrishnan, P.~Priakanth, Gautam Srivastava, and Giancarlo Fortino,
\newblock ``Ewps: Emergency data communication in the internet of medical
  things,''
\newblock {\em IEEE Internet of Things Journal}, vol. 8, no. 14, pp.
  11345--11356, 2021.

\bibitem{isolated-rpm}
Riya Tapwal, Sudip Misra, and Pallav~Kumar Deb,
\newblock ``i-sheet: A low-cost bedsheet sensor for remote diagnosis of
  isolated individuals,''
\newblock {\em IEEE Sensors Journal}, vol. 23, no. 2, pp. 906--913, 2023.

\bibitem{elderly-rpm}
Bin Zhang, Leqi Zhu, Zichen Pei, Qian Zhai, Junhong Zhu, Xiang Zhong, Jingang
  Yi, and Tao Liu,
\newblock ``A framework for remote interaction and management of home care
  elderly adults,''
\newblock {\em IEEE Sensors Journal}, vol. 22, no. 11, pp. 11034--11044, 2022.

\bibitem{smart-rpm}
Eesha Tur~Razia Babar and Mujeeb~U. Rahman,
\newblock ``A smart, low cost, wearable technology for remote patient
  monitoring,''
\newblock {\em IEEE Sensors Journal}, vol. 21, no. 19, pp. 21947--21955, 2021.

\bibitem{cemg-rpm}
Charn~Loong Ng, Mamun Bin~Ibne Reaz, and Muhammad Enamul~Hoque Chowdhury,
\newblock ``A low noise capacitive electromyography monitoring system for
  remote healthcare applications,''
\newblock {\em IEEE Sensors Journal}, vol. 20, no. 6, pp. 3333--3342, 2020.

\bibitem{smart-garment}
Jun Xu, Yucong Zhou, Cheng Zhang, Yitong Li, Xuehui Ma, Dali Ma, and Changyun
  Miao,
\newblock ``Development and evaluation of a respiratory monitoring smart
  garment based on notched optical fiber sensing fabric,''
\newblock {\em IEEE Sensors Journal}, vol. 22, no. 15, pp. 14892--14902, 2022.

\bibitem{optimum}
Avani Vyas and Sujata Pal,
\newblock ``Optimum placement of relay nodes in wbans for improving the qos of
  indoor rpm system,''
\newblock {\em IEEE Sensors Journal}, vol. 21, no. 13, pp. 14434--14442, 2021.

\bibitem{rpm-security}
Mu-Yen Chen,
\newblock ``Establishing a cybersecurity home monitoring system for the
  elderly,''
\newblock {\em IEEE Transactions on Industrial Informatics}, vol. 18, no. 7,
  pp. 4838--4845, 2022.

\bibitem{remote-data}
Charn~Loong Ng, Mamun Bin~Ibne Reaz, and Muhammad Enamul~Hoque Chowdhury,
\newblock ``A low noise capacitive electromyography monitoring system for
  remote healthcare applications,''
\newblock {\em IEEE Sensors Journal}, vol. 20, no. 6, pp. 3333--3342, 2020.

\bibitem{energy-wearable}
Mian~Ahmad Jan, Fazlullah Khan, Spyridon Mastorakis, Muhammad Adil, Aamir
  Akbar, and Nicholas Stergiou,
\newblock ``Lightiot: Lightweight and secure communication for energy-efficient
  iot in health informatics,''
\newblock {\em IEEE Transactions on Green Communications and Networking}, vol.
  5, no. 3, pp. 1202--1211, 2021.

\bibitem{server-overload}
Marah~R. Bataineh, Wail Mardini, Yaser~M. Khamayseh, and Muneer Masadeh~Bani
  Yassein,
\newblock ``Novel and secure blockchain framework for health applications in
  iot,''
\newblock {\em IEEE Access}, vol. 10, pp. 14914--14926, 2022.

\bibitem{service-usability}
Alireza Ghods, Kathleen Caffrey, Beiyu Lin, Kylie Fraga, Roschelle Fritz,
  Maureen Schmitter-Edgecombe, Christopher Hundhausen, and Diane~J. Cook,
\newblock ``Iterative design of visual analytics for a clinician-in-the-loop
  smart home,''
\newblock {\em IEEE Journal of Biomedical and Health Informatics}, vol. 23, no.
  4, pp. 1742--1748, 2019.

\bibitem{service-support}
Sk.~Hasane Ahammad, Md. Zia~Ur Rahman, L.~Koteswara Rao, Asiya Sulthana,
  Navarun Gupta, and Aimé Lay-Ekuakille,
\newblock ``A multi-level sensor-based spinal cord disorder classification
  model for patient wellness and remote monitoring,''
\newblock {\em IEEE Sensors Journal}, vol. 21, no. 13, pp. 14253--14262, 2021.

\bibitem{utilization}
Weizhe Zhang, Rahul Yadav, Yu-Chu Tian, Sumarga Kumar~Sah Tyagi, Ibrahim~A.
  Elgendy, and Omprakash Kaiwartya,
\newblock ``Two-phase industrial manufacturing service management for energy
  efficiency of data centers,''
\newblock {\em IEEE Transactions on Industrial Informatics}, vol. 18, no. 11,
  pp. 7525--7536, 2022.

\bibitem{energy}
Rahul Yadav, Weizhe Zhang, Ibrahim~A. Elgendy, Guozhong Dong, Muhammad Shafiq,
  Asif~Ali Laghari, and Shiv Prakash,
\newblock ``Smart healthcare: Rl-based task offloading scheme for edge-enable
  sensor networks,''
\newblock {\em IEEE Sensors Journal}, vol. 21, no. 22, pp. 24910--24918, 2021.

\bibitem{rigid-flex}
Taiyang Wu, Fan Wu, Chunkai Qiu, Jean-Michel Redouté, and Mehmet~Rasit Yuce,
\newblock ``A rigid-flex wearable health monitoring sensor patch for
  iot-connected healthcare applications,''
\newblock {\em IEEE Internet of Things Journal}, vol. 7, no. 8, pp. 6932--6945,
  2020.

\bibitem{tele-ecg}
Haydar Ozkan, Orhan Ozhan, Yasemin Karadana, Muhammed Gulcu, Samet Macit, and
  Fasahath Husain,
\newblock ``A portable wearable tele-ecg monitoring system,''
\newblock {\em IEEE Transactions on Instrumentation and Measurement}, vol. 69,
  no. 1, pp. 173--182, 2020.

\bibitem{review-wearable}
Shumaila Javaid, Sherali Zeadally, Hamza Fahim, and Bin He,
\newblock ``Medical sensors and their integration in wireless body area
  networks for pervasive healthcare delivery: A review,''
\newblock {\em IEEE Sensors Journal}, vol. 22, no. 5, pp. 3860--3877, 2022.

\bibitem{smart-dfu}
Bill Cassidy, Neil~D. Reeves, Joseph~M. Pappachan, Naseer Ahmad, Samantha
  Haycocks, David Gillespie, and Moi~Hoon Yap,
\newblock ``A cloud-based deep learning framework for remote detection of
  diabetic foot ulcers,''
\newblock {\em IEEE Pervasive Computing}, vol. 21, no. 2, pp. 78--86, 2022.

\bibitem{smart-virtu}
Octavian Postolache, D.~Jude Hemanth, Ricardo Alexandre, Deepak Gupta, Oana
  Geman, and Ashish Khanna,
\newblock ``Remote monitoring of physical rehabilitation of stroke patients
  using iot and virtual reality,''
\newblock {\em IEEE Journal on Selected Areas in Communications}, vol. 39, no.
  2, pp. 562--573, 2021.

\bibitem{fog-cloud}
Forough Shirin~Abkenar and Abbas Jamalipour,
\newblock ``Energy optimization in association-free fog-iot networks,''
\newblock {\em IEEE Transactions on Green Communications and Networking}, vol.
  4, no. 2, pp. 404--412, 2020.

\bibitem{fog-offload2}
Rahul Yadav, Weizhe Zhang, Omprakash Kaiwartya, Houbing Song, and Shui Yu,
\newblock ``Energy-latency tradeoff for dynamic computation offloading in
  vehicular fog computing,''
\newblock {\em IEEE Transactions on Vehicular Technology}, vol. 69, no. 12, pp.
  14198--14211, 2020.

\bibitem{fog-offload}
Junyu Ren, Jinze Li, Huaxing Liu, and Tuanfa Qin,
\newblock ``Task offloading strategy with emergency handling and blockchain
  security in sdn-empowered and fog-assisted healthcare iot,''
\newblock {\em Tsinghua Science and Technology}, vol. 27, no. 4, pp. 760--776,
  2022.

\bibitem{cnn-sound}
Chien-Yao Wang, Tzu-Chiang Tai, Jia-Ching Wang, Andri Santoso, Seksan
  Mathulaprangsan, Chin-Chin Chiang, and Chung-Hsien Wu,
\newblock ``Sound events recognition and retrieval using
  multi-convolutional-channel sparse coding convolutional neural networks,''
\newblock {\em IEEE/ACM Transactions on Audio, Speech, and Language
  Processing}, vol. 28, pp. 1875--1887, 2020.

\bibitem{baucas}
Marc~Jayson Baucas and Petros Spachos,
\newblock ``Speech recognition driven assistive framework for remote patient
  monitoring,''
\newblock in {\em 2019 IEEE Global Conference on Signal and Information
  Processing (GlobalSIP)}, 2019, pp. 1--5.

\bibitem{warden2018speech}
Pete Warden,
\newblock ``Speech commands: A dataset for limited-vocabulary speech
  recognition,'' 2018.

\end{thebibliography}
\end{document}